\pdfoutput=1

\documentclass[twocolumn]{aastex6}

\usepackage{graphicx, latexsym, textcomp, gensymb, multirow,booktabs, amsfonts,amsmath,amssymb, natbib, url, hyperref}
\usepackage[space]{grffile}

\newif\iflatexml\latexmlfalse
\DeclareGraphicsExtensions{.pdf,.PDF,.png,.PNG,.jpg,.JPG,.jpeg,.JPEG}

\usepackage[utf8]{inputenc}
\usepackage[english]{babel}

\newcommand{\HI}{H{\sc\ i}}
\newcommand{\HII}{H{\sc\ ii}}
\newcommand{\OI}{O{\sc\ i}}
\newcommand{\SiII}{Si{\sc\ ii}}
\newcommand{\SiIII}{Si{\sc\ iii}}
\newcommand{\CII}{C{\sc\ ii}}
\newcommand{\FeII}{Fe{\sc\ ii}}

\newcommand{\SII}{S{\sc\ ii}}
\newcommand{\SiIV}{Si{\sc\ iv}}
\newcommand{\PII}{P{\sc\ ii}}
\newcommand{\MgII}{Mg{\sc\ ii}}
\newcommand{\NaI}{Na{\sc\ i}}
\newcommand{\kms}{km ${\rm s^{-1}}$}

\shorttitle{An Ionized Very-High Velocity Cloud Toward M33}
\shortauthors{Y. Zheng et al. }

\begin{document}

\defcitealias{Fox14}{F14}
\defcitealias{Wright79}{W79}
\defcitealias{Zheng17}{Z17}
\defcitealias{Chisholm16}{C16}


\title{The Discovery and Origin of A Very-High Velocity Cloud Toward M33}

\author{Y. Zheng$^1$, J. K. Werk$^2$,, J. E. G. Peek$^3$,  M. E. Putman$^1$}
\affil{$^1$ Department of Astronomy, Columbia University, New York, NY 10027, USA; yzheng@astro.columbia.edu \\
       $^2$ Department of Astronomy, University of Washington, Seattle, WA 98195-1580, USA \\
       $^3$ Space Telescope Science Institute, 3700 San Martin Dr, Baltimore, MD 21218, USA }

\begin{abstract}
We report the detection of a largely ionized very-high velocity cloud (VHVC; $v_{\rm LSR}\sim-350$ \kms) toward M33 with the  {\it Hubble Space Telescope}/Cosmic Origin Spectrograph. The VHVC is detected in \OI, \CII, \SiII, and \SiIII\ absorption along five sightlines separated by $\sim0.06-0.4\degree$. On sub-degree scales, the velocities and ionic column densities of the VHVC remain relatively smooth with standard deviations of $\pm$14 \kms\ and $\pm$0.15 dex between the sightlines, respectively. The VHVC has a metallicity of [\OI/\HI]$=-0.56\pm0.17$ dex (Z$=0.28\pm0.11$ Z$_{\odot}$). Despite the position-velocity proximity of the VHVC to the ionized Magellanic Stream, the VHVC's higher metallicity makes it unlikely to be associated with the Stream, highlighting the complex velocity structure of this region of sky. We investigate the VHVC's possible origin by revisiting its surrounding \HI\ environment. We find that the VHVC may be: (1) a MW CGM cloud, (2) related to a nearby \HI\ VHVC -- Wright's Cloud, or (3) connected to M33's northern warp. Furthermore, the VHVC could be a bridge connecting Wright's Cloud and M33's northern warp, which would make it a Magellanic-like structure in the halo of M33. 
\end{abstract}


\section{Introduction}
\label{sec1}

The inner halo ($\lesssim20$ kpc) of the Milky Way (MW) is populated by neutral \HI\  high-velocity clouds (HVCs) and their ionized counterparts (ionized HVCs). HVCs are usually defined to have velocities of $|v_{\rm LSR}|$$>$90 \kms, where $v_{\rm LSR}$ is the velocity in the rest frame of the Local Standard-of-Rest (LSR) with respect to the Sun. The neutral and ionized HVCs exist at an interface where outflows from the Galactic disk is likely to be closely interacting with inflows from the halo (see \citealt{Putman12} and references therein). HVCs are commonly arranged in large complexes observed in \HI\ 21-cm emission \citep{Wakker91};  on the other hand, ionized HVCs are detected via optical or ultraviolet (UV) absorption lines toward background stars or quasars (see \citealt{vanWoerden04} and references therein). The metallicities of HVCs vary from $\sim$0.1 solar (e.g., complex A, WD), to $\sim$0.1$-$0.3 solar (e.g., complex C; \citealt{Tripp03, Shull11}) and likely near solar (e.g., complex M, WB), complicating the interpretation of the origins \citep{vanWoerden04}. Distances to HVCs are difficult to obtain; the only direct means is to bracket the distance via the detection (non-detection) of absorption lines at similar velocities toward background (foreground) stars projected along the line of sight to the clouds (e.g, \citealt{Thom06, Thom08, Lehner10, Smoker11, Peek16}). So far, it is generally suggested that most of the clouds lie within 5$-$15 kpc of the Sun (e.g., \citealt{Wakker01, Wakker08, Lehner12}).

\begin{figure*}[t]
\begin{center}
\includegraphics[width=\textwidth]{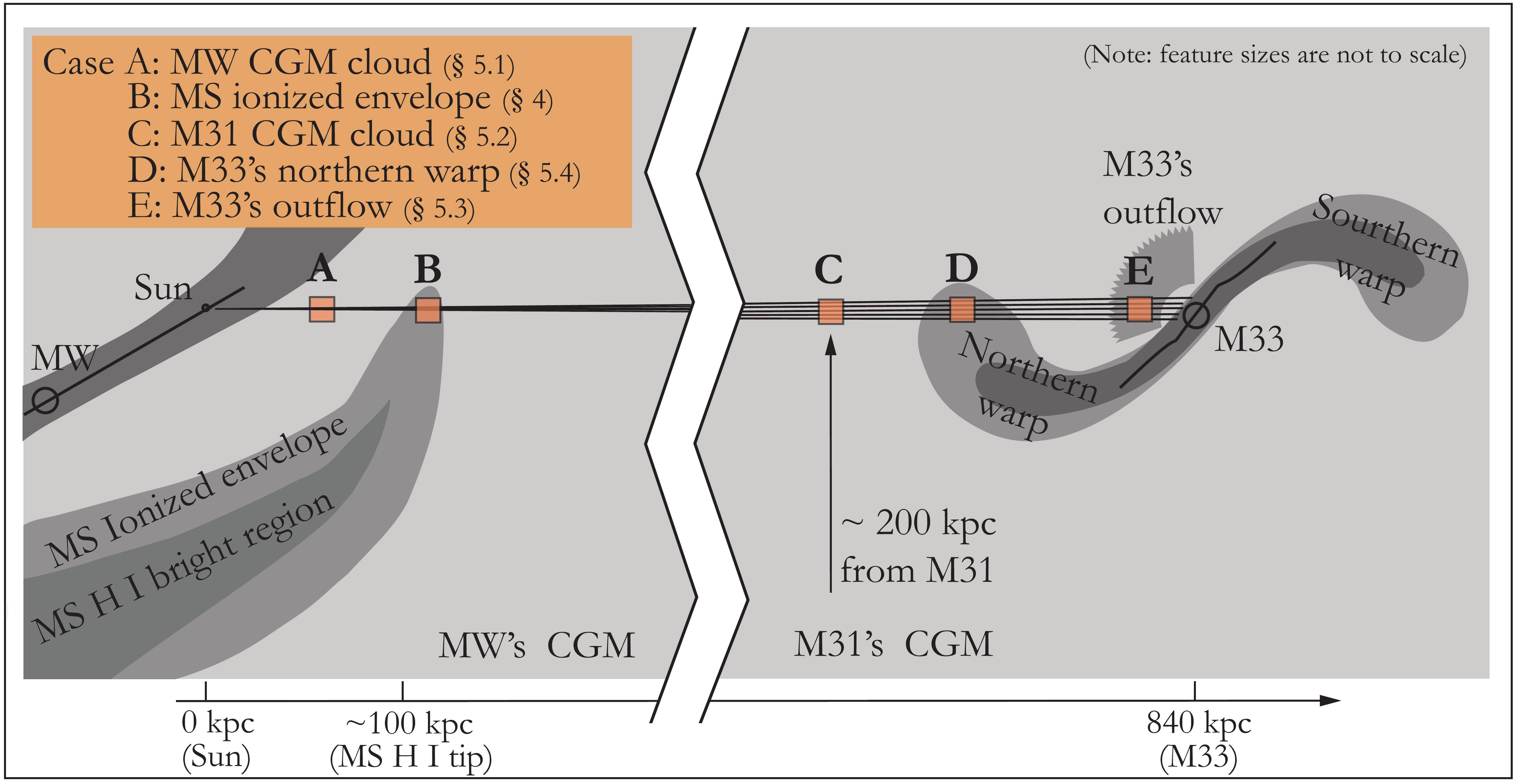}
\caption{A cartoon showing the potential locations (A/B/C/D/E) of the VHVC along the {\it HST}/COS sightlines. Feature sizes are not to scale and the sizes of the orange squares at different locations are arbitrary. \HI\ features are shown in dark gray while ionized features are indicated in light gray. We will discuss Case B -- the ionized envelope of the Magellanic Stream in Section \ref{sec4}, and the other cases in Section \ref{sec5}. }
\label{fig1}
\end{center}
\end{figure*}

One \HI\ complex that does not lie in MW's inner halo is the Magellanic Stream, a $\sim$200$\degree$ long \HI\ structure that originates from the Magellanic Clouds (MCs; \citealt{Putman03, Nidever08, Nidever10}). The Stream has been stripped from the MCs by tides and/or ram pressure forces as the two galaxies move through the MW halo (e.g., \citealt{Murai80, Gardiner94, Besla10}). The closest distance of the Stream is likely $\sim$55 kpc near the MCs while the most distant point is suggested to be at $\sim$100 kpc near the tail based on tidal models \citep{Besla12} and geometrodynamical calculation \citep{Jin08}. Except very close to the MCs, the metallicity of the Stream is $\sim$0.1 solar, measured from several QSO sightlines that pass through different locations of the Stream \citep{Fox10, Fox13, Richter13, Gibson00}. Near the Stream, numerous small individual \HI\ HVCs are found moving at Magellanic-like velocities \citep{Braun04, Bruns05, Stanimirovic08, Nidever10}. In addition, \citet[][hereafter F14]{Fox14} found that the Magellanic System, including the Stream, the Bridge and the Leading Arm, is likely to be surrounded by an ionized extended envelope with a cross-section of $\approx$$11, 000$ deg$^2$, which is nearly four times larger that its \HI-bright region. Excluding the MCs themselves, the Magellanic \HI-bright regions\footnote{Including the Magellanic Stream, the Magellanic Bridge, the Leading Arm, see \citetalias{Fox14}.} have total mass of M(\HI+\HII)$\sim$$1.4\times10^9(d/55\ {\rm kpc})^2$ M$_{\odot}$ (\citealt{Bruns05}; \citetalias{Fox14}), and its ionized envelope beyond the main body has an ionized mass of M(\HII)$\sim$$5.5\times10^8(d/55\ {\rm kpc})^2$ M$_{\odot}$ \citepalias{Fox14}. 

On the projected sky, the ionized envelope of the Magellanic Stream near its tip may overlap with M31's and M33's circumgalactic medium (CGM) seen toward extragalactic sources (e.g., quasars; QSOs) and in emission, complicating the interpretation of the gas velocities in this region (\citetalias{Fox14}; \citealt{Nidever10, Lehner15}). The tip of the Stream and its surrounding \HI\ clouds move at $-$500$\lesssim$$v_{\rm LSR}$$\lesssim$$-$300 \kms\ (\citetalias{Fox14}; \citealt{Nidever10, Lehner15}). Since M31's and M33's systemic velocities are $-$301 \kms\ \citep{Courteau99} and $-$179 \kms\ \citep{Corbelli97} respectively, it is possible that the CGM of these two galaxies may have velocities comparable to the Magellanic Stream \citep{Lockman12, Lehner15, Wolfe16}. In addition, this region is close to the area studied in \cite{Richter16} who suggested nearby absorbers are likely to be streaming toward the Local Group center.

\begin{figure*}[t]
\begin{center}
\includegraphics[width=\textwidth]{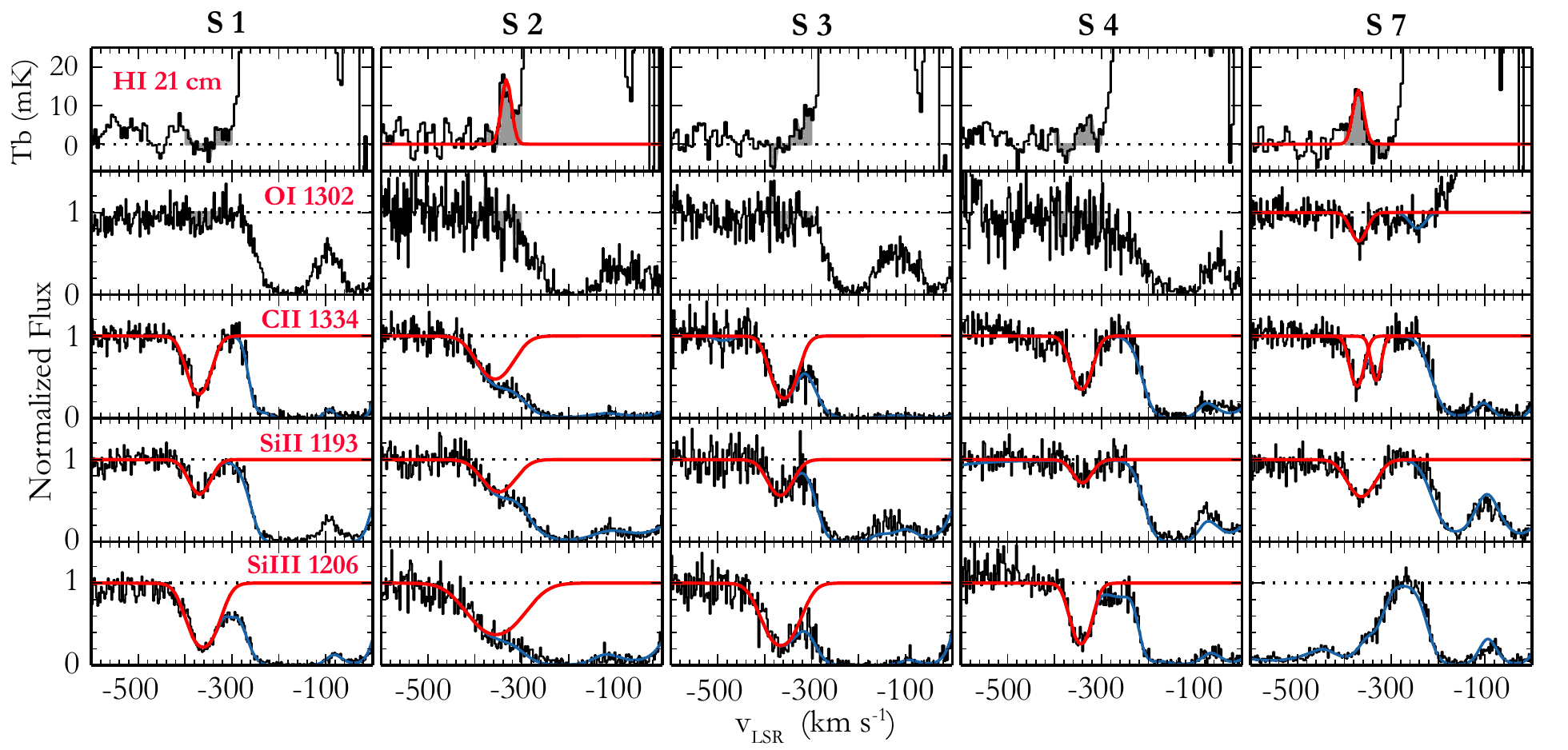}
\caption{The top row shows the \HI\ 21cm spectra from AGES; the red curves are Gaussian fits to the two possible \HI\ detections with $>$2$\sigma$ significance. The data have a brightness temperature sensitivity of 4-8 mK \citep{Auld06}. The bottom four rows show the continuum-normalized \OI, \CII, \SiII, and \SiIII\ absorption lines and their Voigt-profile fits (red); the underlying blue curves indicate the overall fits. The absorption components at $v_{\rm LSR}\sim-350$ \kms\ are associated with the VHVC, and those at $-300<v_{\rm LSR}<-100$ \kms\ ($\sim$0 \kms) are from M33 (MW). }
\label{fig2}
\end{center}
\end{figure*}

To intensify the intrigue of this Magellanic-M33-M31 overlap region, here we report the detection of a VHVC along five sightlines toward M33, and therefore the VHVC lies in front of M33's disk. As shown in Fig \ref{fig1}, between M33 and MW,  there are five possibilities that may cause the absorption: A -- a MW CGM cloud, B -- the ionized envelope of the Magellanic Stream, C -- an M31 CGM cloud, D -- M33's northern warp, and E -- M33's outflow. In addition, on the projected sky the VHVC is $\sim$2$\degree$ from an \HI-VHVC -- Wright's Cloud (\citealt{Wright74}; \citealt[][hereafter, W79]{Wright79}) and they are at similar velocities.  The distance to Wright's Cloud is not clear yet; we discuss the relation of the VHVC and Wright's Cloud and the implication in Section \ref{sec5.5}. 

In the following, we present our data and discuss the origin possibilities of the VHVC and its possible connections with nearby \HI\ structures. In Section \ref{sec2} we show absorption-line data obtained from {\it HST}/COS, and \HI\ 21-cm data from the Arecibo Galaxy Environment Survey (AGES; \citealt{Keenan16}) and the Galactic Arecibo L-Band Feed Array \HI\ survey (GALFA-\HI; \citealt{Peek11}). In Section \ref{sec3}, we calculate the \OI\ abundance and show the gas properties over sub-degree scale for the VHVC. In Section \ref{sec4}, we discuss the inferences that can be drawn from the metallicity and the spatial patchiness of the VHVC in relation to the Magellanic Stream (Case B in Fig \ref{fig1}). We discuss the possible origins (Case A, C, D, and E in Fig \ref{fig1}) of the VHVC in Section \ref{sec5} and summarize our findings in Section \ref{sec6}.

\section{Data and Measurement}
\label{sec2}

\subsection{HST/COS Observations}
\label{sec2.1}

The VHVC is a serendipitous detection in the observations described in \citet[][hereafter Z17]{Zheng17}. Among the seven sightlines (S1-S7) that were used in \citetalias{Zheng17}, five show significant ion absorption lines at $v_{\rm LSR}\sim-350$ \kms, while the other two are heavily contaminated by stellar lines and/or M33's interstellar absorption lines at the corresponding velocities. Therefore, in this work we only use the five sightlines, which are M33-UIT-236, M33-FUV-350, M33-FUV-444, NGC 592, and M33-FUV-016. Hereafter we use their sightline ID S1, S2, S3, S4, S7, same as those in \citetalias{Zheng17} for consistency. 

The VHVC is detected in absorption lines of \OI\ $\lambda$1302, \CII\ $\lambda$1334, \SiII\ $\lambda\lambda$1190, 1193\footnote{We do not use \SiII\ $\lambda$ 1260 \AA\ because it is blended with \SII\ $\lambda$ 1259 \AA\ absorption from M33 (see \citetalias{Zheng17}).}, and \SiIII\ $\lambda$1206. Hereafter we refer to these ions as ``low ions" since they represent low ionization states that require $<$20 eV to be produced\footnote{\SiII/\SiIII/\CII\ need 8.15/16.35/11.26 eV to be ionized from their previous states, respectively.}. We do not detect \FeII, \SII, \PII, \SiIV\ absorption lines at similar velocities even though our data cover these ionic species; the lack of these ions may reflect some limits on metallicity and ionization condition, which we decide not to explore further since it requires sophisticated ionization assumptions and modelings. 

The {\it HST}/COS observations have been thoroughly described in \citetalias{Zheng17}, here we briefly summarize relevant details. The spectra were obtained using the G130M FUV grating with a central wavelength of 1291 \AA\ and a wavelength coverage of 1134$-$1431 \AA. The velocity resolution (FWHM) is 14$-$19 \kms\ per resolution element (COS data handbook; \citealt{Fox15}). The spectra were taken with the Primary Science Aperature which has an aperture size of 2.5 arcseconds. We retrieved the calibrated and co-added data from the Mikulski Archive from Space Telescopes (MAST). The spectra have been processed by the standard CalCOS pipeline (version 3.0). In \citetalias{Zheng17}, we have justified the CalCOS products are reliable for our scientific analysis, mainly benefiting from the simple setup of our observations. The COS spectra used in this work can be found here: \href{http://dx.doi.org/10.17909/T9FG6R}{10.17909/T9FG6R}. 

Among the detected ionic absorption lines, those of \OI\ 1302 \AA\ require special treatment. From the original CalCOS co-added spectra, we find a clear \OI\ absorption line along S7. The other four spectra are dominated by strong air-glow emission at the expected velocities due to \OI\ in the exosphere of the Earth. To confirm the S7 detection and to check if other sightlines also show similar \OI\ signals, we conducted night-only data reduction following the night-only calibration tutorial\footnote{\href{url}{https://justincely.github.io/AAS224/timefilter{\_}tutorial.html}} provided by Justin Ely. Briefly, we installed the CalCOS v3.0 pipeline locally and retrieved new calibration reference files using the {\it HST} Calibration Reference Data System (CRDS). We used the TimeFilter python module to select the night-only photons from the raw data, then constructed the \OI\ night-only spectra for the five sightlines. From the night-only spectra, we find that S7 presents the same \OI\ $\lambda$1302 absorption as seen in the original spectrum but with a noisier profile due to the reduced number of photons. However, S1-S4 still do not show \OI\ absorption, which is most likely due to the low S/N ratio of the night-only spectra.

With the reduced spectra, we perform continuum and Voigt-profile fitting following the procedures outlined in \citetalias{Zheng17}. Briefly, for singular transition we normalize the absorption profile by fitting stellar continuum and stellar lines with Legendre polynominals within $\pm$1000 \kms\ from the line center. The fitting is evaluated using the reduced $\chi^2$ values. For the Voigt-profile fitting, we fit one component for each line unless obvious multiple components exist. The Voigt-profile fits give estimates for the centroid velocity $v$, Doppler width $b$, and column density log $N_{\rm ion}$, and the fitting is performed multiple times iteratively so that these parameters converge. In Fig \ref{fig2}, we show the \OI, \CII, \SiII, and \SiIII\ absorption lines which have been continuum-normalized and Voigt-profile fitted, and in Table \ref{tb1} we tabulate the best-fit $v$, log $N$, and FWHM($\equiv1.667b$). Overall, all lines yield straightforward solutions except those detected along S2, which has considerable larger uncertainties due to the blending with M33's interstellar absorption at less negative velocities. In addition, for \OI\ absorption lines in Fig \ref{fig2}, we use the night-only spectra for S1-S4 since their original spectra at $v_{\rm LSR}\sim-350$ \kms\ are dominated by geocoronal airglow emission. For S7, we use the original CalCOS co-added spectrum because the absorption is isolated from the airglow emission at $v_{\rm LSR}>-200$ \kms\ and this spectrum has a higher S/N than the night-only.

\begin{table} 
\renewcommand{\arraystretch}{0.95}
\begin{center}
\caption{{Ion Properties}}
    \begin{tabular}{ c c c c}
        \hline
        \hline
        Sightline$^a$ & v$^b$ & log N$^c$ & FWHM$^d$  \\ 
                      & \kms & cm$^{-2}$ & \kms \\
        \hline
        \hline
        \multicolumn{4}{c}{\HI}\\
        \hline
        S1 & $[-400, -300]$       & $<$17.68       & -     \\ 
        S2 & $-332.7$             & 17.91$\pm$0.13 & 25.3  \\
        S3 & $[-400, -300]$       & $<$17.94       & -     \\
        S4 & $[-400, -300]$       & $<$17.92       & -     \\
        S7 & $-370.3$             & 17.87$\pm$0.14 & 27.0  \\
        \hline
        Mean$^e$ & $-351.5$$\pm$18.8 & 17.89$\pm$0.02 & 26.1$\pm$0.8 \\ 
        \hline
        \hline
        \multicolumn{4}{c}{\OI}\\
        \hline
        S1   & [$-$400, $-$300] & $<$14.29 & - \\
        S2   & [$-$400, $-$300] & $<$14.43 & - \\
        S3   & [$-$400, $-$300] & $<$14.18 & - \\
        S4   & [$-$400, $-$300] & $<$14.38 & - \\
        S7   & $-$354.1$\pm$2.7 & 14.00$\pm$0.07 & 33.5 \\
        \hline
        Mean & $-$354.1 & 14.00 & 33.5 \\
        \hline
        \hline
        \multicolumn{4}{c}{\SiII} \\
        \hline
        S1   & $-$354.9$\pm$1.8 & 13.26$\pm$0.01 & 47.8 \\
        S2   & $-$334.6$\pm$6.8 & 13.41$\pm$0.09 & 74.0 \\
        S3   & $-$353.2$\pm$3.9 & 13.35$\pm$0.06 & 55.3 \\
        S4   & $-$325.8$\pm$3.4 & 13.01$\pm$0.07 & 42.8 \\
        S7   & $-$348.5$\pm$2.1 & 13.40$\pm$0.03 & 60.5 \\
        \hline
        Mean & $-$343.4$\pm$11.3 & 13.29$\pm$0.15 & 56.1$\pm$10.8 \\
        \hline
        \hline
        \multicolumn{4}{c}{\SiIII} \\
        \hline
        S1   & $-$347.5$\pm$1.2 & 13.29$\pm$0.02 & 62.5 \\
        S2   & $-$340.8$\pm$6.4 & 13.37$\pm$0.04 & 116.7 \\
        S3   & $-$351.8$\pm$5.5 & 13.33$\pm$0.06 & 73.2 \\
        S4   & $-$328.6$\pm$1.9 & 13.09$\pm$0.04 & 42.5 \\
        S7   & -$^{f}$ & - & - \\
        \hline
        Mean & $-$342.2$\pm$8.8 & 13.27$\pm$0.11 & 73.7$\pm$27.1 \\
        \hline
        \hline
        \multicolumn{4}{c}{\CII}  \\
        \hline
        S1   & $-$356.6$\pm$1.2 & 14.17$\pm$0.02 & 48.7 \\
        S2   & $-$342.5$\pm$6.5 & 14.14$\pm$0.10 & 78.0 \\
        S3   & $-$345.1$\pm$2.3 & 14.26$\pm$0.04 & 52.8 \\
        S4   & $-$327.4$\pm$1.5 & 14.05$\pm$0.03 & 43.3 \\
        S7   & $-$357.7$\pm$1.5 & 13.85$\pm$0.04 & 29.0 \\
             & $-$316.8$\pm$1.5 & 13.63$\pm$0.06 & 19.8 \\
        \hline
        Mean & $-$341.8$\pm$9.6 & 14.13$\pm$0.08 & 54.3$\pm$12.2 \\
        \hline
        \hline
    \end{tabular} 
\end{center}
    \tablecomments{{\footnotesize ($a$) S1: M33-UIT-236; S2: M33-FUV-350; S3: M33-FUV-444; S4: NGC 592; S7: M33-FUV-016. See Table 1 in \citetalias{Zheng17} for more information. ($b$): the Gaussian (Voigt-profile) fitted centroid velocities for \HI\ (\OI, \CII, \SiII, and \SiIII). ($c$): Gaussian (Voigt-profile) fitted logarithm column densities; if there is no detection a 3$\sigma$ upper limit is indicated. ($d$): For \HI, FWHM$\equiv2.355\sigma$ where $\sigma$ is from Gaussian fits. For \OI, \CII, \SiII, and \SiIII, FWHM$\equiv1.667b$ where $b$ is the Doppler width values from Voigt-profile fits. ($e$): the mean and the standard deviation of the measured values. For C $\rm \scriptstyle II$ along S7, the two components are considered as one while calculating the overall mean and standard deviation. ($f$): \SiIII\ along S7 is contaminated by strong stellar lines, so we do not use it. }}
\label{tb1}
\end{table}

\subsection{\HI\ 21-cm data}
\label{sec2.2}

The \HI\ 21-cm spectra shown in the top row of Fig \ref{fig2} are from the Arecibo Galaxy Environment Survey (AGES; \citealt{Keenan16}). AGES data have an angular resolution of $\sim$4', spectral resolution of $\sim$5.2 \kms, and a 1$\sigma$ sensitivity of $\sim$1.5$\times$10$^{17}$ cm$^{-2}$ over 10 \kms. Only S2 and S7 indicate $>$2$\sigma$ emission signals, which we fit with Gaussian components. S1, S3, and S4 do not show significant detection above the noise level, thus we integrate the spectra from $-400$ to $-300$ \kms\ to derive a 3$\sigma$ upper limit. Note that this $[-400, -300]$ \kms\ integration range is relatively broad as compared to the narrow \HI\ lines detected in S2 and S7; the derived 3$\sigma$ upper limits should be conservative estimates for the \HI\ along S1, S3, and S4. We tabulate the Gaussian-fitted \HI\ velocities and column densities and the integrated upper limits in Table \ref{tb1}; the column density values are calculated using N(\HI)$=1.823\times10^{18}\int^{-300}_{-400} T_{\rm B}\ dv$ cm$^{-2}$ \citep{Draine11}. Note that \HI-bearing gas may not be entirely co-spatial with the VHVC detected along {\it HST}/COS sightlines, since AGES beam size ($\sim$4') is $\sim$100 times larger than the COS aperture size (2.5"). Still, the high-sensitivty AGES data provide the best available estimate of the \HI\ for the VHVC.

In addition, we use \HI\ 21-cm data from the Galactic Arecibo L-Band Feed Array \HI\ survey (GALFA-\HI; \citealt{Peek11}). The GALFA-\HI\ data is from its second data release (Peek et al. 2017, in prep.), with an angular resolution of $\sim$4', a spectral resolution of $\sim$0.18 \kms, and a 1$\sigma$ noise level of 140 mK in an integrated 1 \kms\ velocity channel. We use the GALFA-\HI\ data to create maps to study \HI\ features near the VHVC in Section \ref{sec5.5}.

\section{Gas Properties of the VHVC}
\label{sec3}

\subsection{Metallicity}
\label{sec3.1}

The \OI\ detection along S7 provides us a direct metallicity measurement for the VHVC without assuming any ionization correction. This is because \OI\ and \HI\ have comparable ionization potentials from their ground states, and \OI\ is not likely to be heavily depleted onto dust \citep{Jenkins09}. As shown in Table \ref{tb1}, we find log N(\OI)$=$14.00$\pm$0.07 cm$^{-2}$ and log N(\HI)$=$17.87$\pm$0.14 cm$^{-2}$ for S7. Thus the gas-phase abundance is 12+log(O/H)$=$8.13$\pm$0.16. Using the oxygen solar abundance (12+log(O/H)$_{\odot}=8.69\pm0.05$; \citealt{Asplund09}), we measure $\rm [O\ {\scriptstyle I}/H\ {\scriptstyle I}] \equiv log(O/H)-log(O/H)_{\odot}=-0.56\pm0.17$ dex, corresponding to a metallicity of Z$\equiv$10$^{\rm [OI/HI]}=$0.28$\pm$0.11 Z$_{\odot}$ for the VHVC.  In addition, sightline S2 shows similar \HI\ emission although \OI\ only has a 3 $\sigma$ upper limit. For S2, we found a 3$\sigma$ upper limit of $\rm [O\ {\scriptstyle I}/H\ {\scriptstyle I}]$$\leq$$-$0.17 dex and Z$\leq$0.68 Z$_{\odot}$, consistent with the value estimated from S7.

\begin{figure}[t]
\begin{center}
\includegraphics[width=\columnwidth]{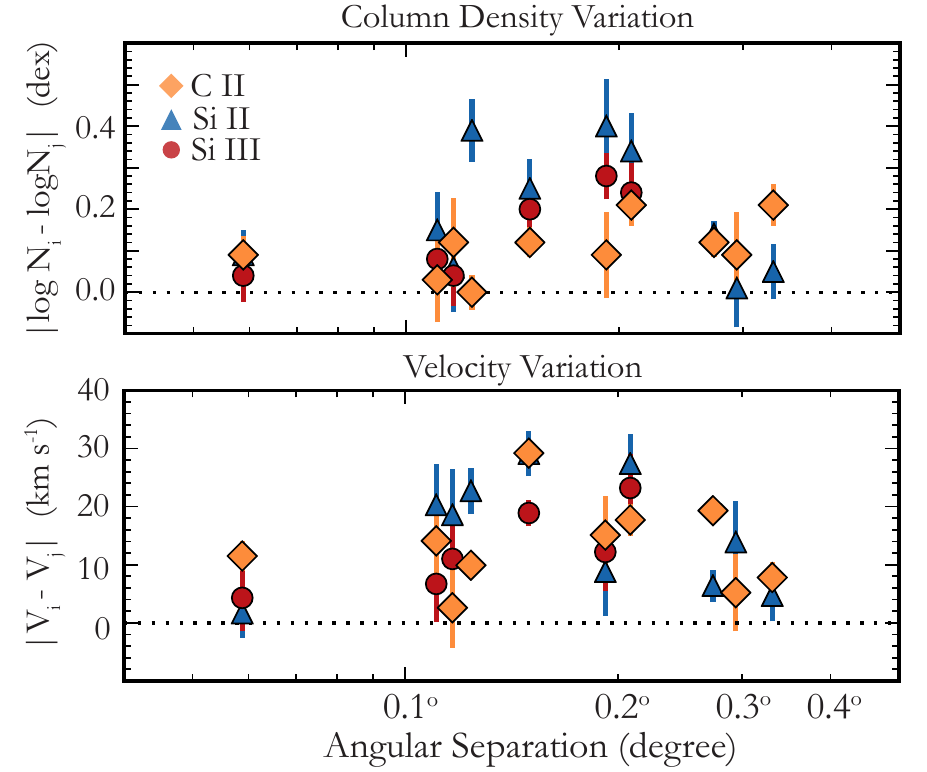}
\caption{The absolute values of ion column density differences (top) and centroid velocity differences (bottom) of paired sightlines over sub-degree scales. \CII/\SiII/SiIII\ are shown as orange diamonds/blue triangles/red circles, respectively. Data points without error bars mean the errors are smaller than the sizes of the symbols. }
\label{fig3}
\end{center}
\end{figure}

\subsection{Sub-Degree Scale Variation}  
\label{sec3.2} 

For each ion, we calculate the mean and standard deviation values of $v$, log N, and FWHM among the five sightlines, which are tabulated in Table \ref{tb1}. Both \HI\ and the four ions show similar mean centroid velocities at  $v_{\rm LSR}\sim$$-$350 \kms\ with a standard deviation of $\sim$15 \kms. Column densities vary up to 0.15 dex as measured from \SiII; in most cases the differences among the sightlines remain $\lesssim$0.1 dex. As for the FWHM, a relatively large standard deviation of 27 \kms\ is seen in \SiIII, contributed by the blended \SiIII\ absorption line along S2. Otherwise the FWHM values of all the transition lines remain $\lesssim$70 \kms.

The above standard deviations are measured among five sightlines with projected angular separations of $\sim$0.06$-$0.4$\degree$. To determine how the absorption properties fluctuate over this sub-degree scale, we group the five sightlines into pairs and calculate the absolute values of the differences:  $|$log N$_i-$log N$_j|$, and $|v_i-v_j|$. From the top panel of Fig \ref{fig3}, we find that the maximum log N difference is $\sim$0.4 dex with an average of $\sim$0.15 dex. The bottom panel shows that the centroid velocity differences between sightlines vary up to 30 \kms\ with an average of $\sim$14 \kms, which is well within COS velocity resolution. The small sightline-to-sightline variations suggest that the VHVC is relatively smooth on sub-degree scales. To better interpret the smooth distribution,  it is necessary to know the physical separations between the five sightlines. We will investigate this in Section \ref{sec5} where the potential origins of the VHVC are discussed. 

\begin{figure}[t]
\begin{center}
\includegraphics[width=\columnwidth]{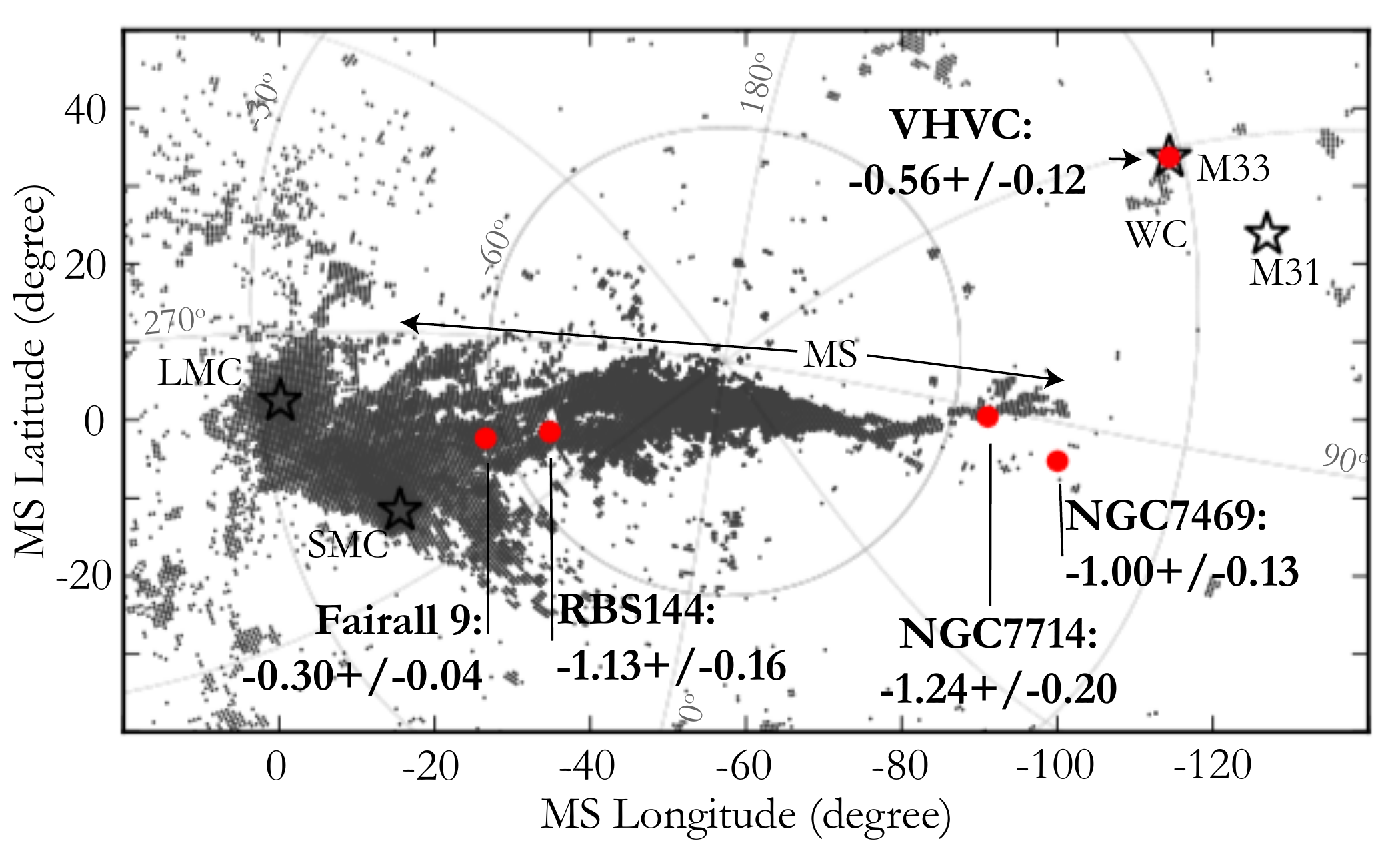}
\caption{The positions of the VHVC, M33, M31, Wright's Cloud (WC), the MCs (LMC, SMC), the Magellanic Stream (MS) trailing the MCs, and four QSO sightlines (Fairall 9, RBS 144, NGC 7714, NGC 7469) with metallicity estimates for the Magellanic Stream. Note that the VHVC is at the same position as M33 and in front of the galaxy. The gray patches are \HI\ data from \cite{Nidever08}, which include more structures besides the Magellanic Stream. The map is in the Magellanic coordinate system \citep{Nidever08}.}  
\label{fig4}%
\end{center}
\end{figure}

\begin{figure}[t]
\begin{center}
\includegraphics[width=\columnwidth]{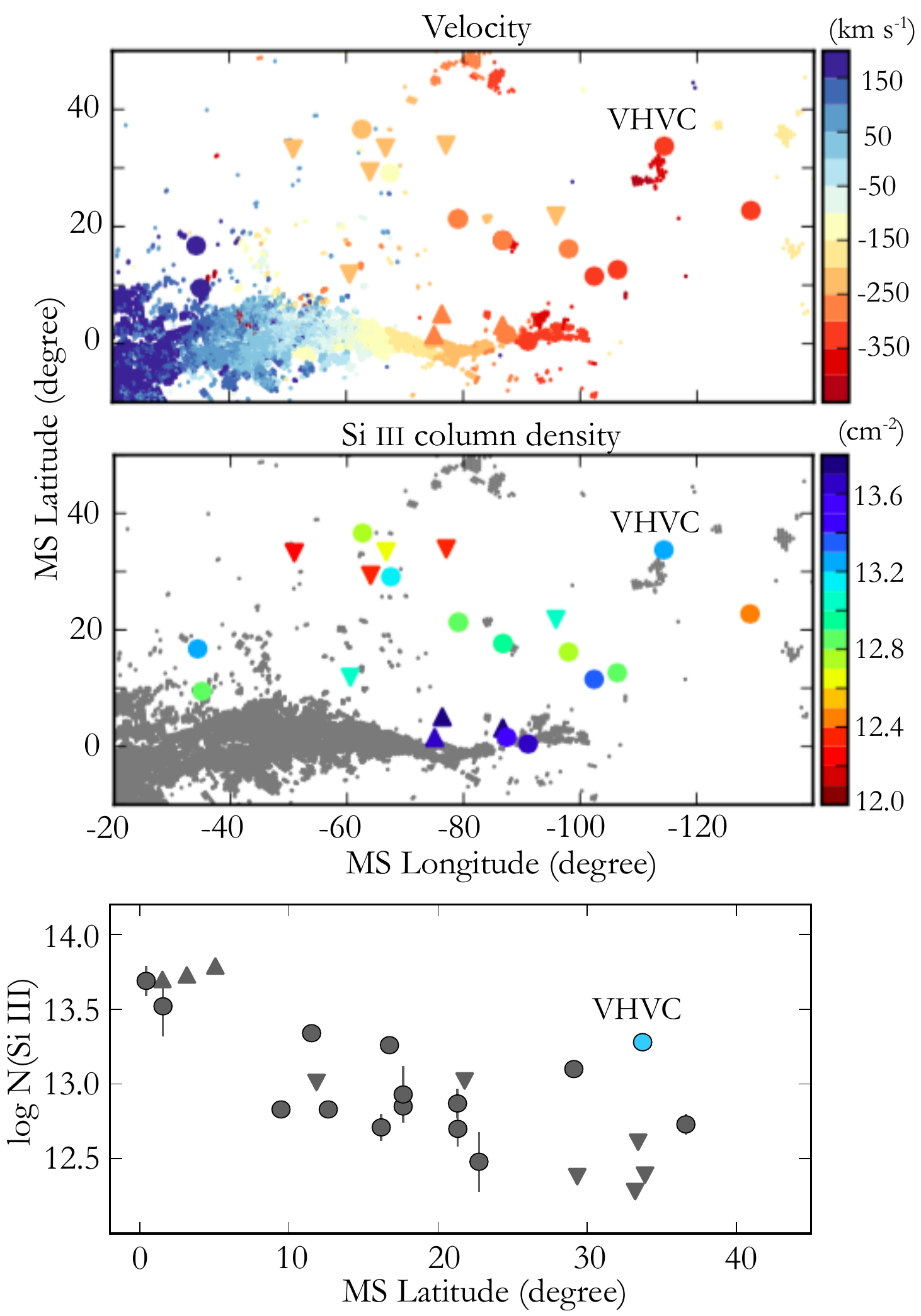}
\caption{Gas velocities (top) and column densities (middle; bottom) near the tip of the Magellanic Stream. The ionized gas absorbers, for which we use \SiIII\ as a representative transition, are detected through background QSOs and indicated by circles (good measurements), upward (lower limits) and downward (upper limits) triangles. All the spectra were taken with COS/G130M, so there is no systematic difference due to instrumental difference in comparing the ion column densities. The Magellanic-associated gas absorber measurements are from \citetalias{Fox14} and the VHVC measurement is from this work. We show the \HI\ tip of the Magellanic Stream; the \HI\ velocities and column densities are from \cite{Nidever08}.  In the bottom panel, data points without error bars mean the errors are smaller than the sizes of the symbols. }
\label{fig5}
\end{center}
\end{figure}

\section{Is the VHVC Related to the Magellanic Stream?}
\label{sec4}

In the vicinity of the Magellanic System, large amounts of ionized gas have been found to move at expected Magellanic velocities (\citetalias{Fox14}; \citealt{Nidever08}). The expected velocities are defined by extrapolating from the \HI\ velocity fields at certain longitudes in the Magellanic coordinate system. Based on this position-velocity proximity, \citetalias{Fox14} suggested the nearby ionized gas is physically associated with the Magellanic System\footnote{Including the Magellanic Stream, the Magellanic Bridge, and the Leading Arm; see \citetalias{Fox14}. }, forming an ionized envelope surrounding the System. The ionized envelope has a total mass of M(\HII)$\sim$5.5$\times$10$^8$(d/55 kpc)$^2$ M$_{\odot}$, contributing $\sim$20\% of the total mass to the System (excluding the MCs themselves). It extends $\approx$$11, 000$ deg$^2$ on the sky, including the region where we find the VHVC, which is $\sim$40$\degree$ from the tip of the Magellanic Stream, as is shown in Fig \ref{fig4} and Case B in Fig \ref{fig1} (see also \citealt{Richter16} for a even larger estimate on the cross-section for the Magellanic System). Thus, this close projected distance naturally leads us to examine whether the VHVC originates from the Magellanic Stream.

Our main argument against a Magellanic origin is based on the single metallicity we measured from the COS spectrum of S7. The VHVC has a metallicity of $\rm [O\ {\scriptstyle I}/H\ {\scriptstyle I}]=-0.56\pm0.17$ dex (Z$=$0.28$\pm$0.11 Z$_{\odot}$; see Section \ref{sec3.1}). As for the Magellanic Stream, its metallicity has been measured at four locations (see Fig \ref{fig4}). \cite{Richter13} found that the Stream close to the MCs has $\rm [S/H]=$$-$0.30$\pm$0.04 (0.50 Z$_{\odot}$; see also \citealt{Gibson00}) toward Fairall 9. This material is likely to be recently stripped metal-rich gas that originated from the LMC. Except at the Fairall 9 position, the main body of the Stream has a mean metallicity of [X/H]$=$$-$1.12 dex,  as calculated at three other positions: $\rm [S\ {\scriptstyle II}/H\ {\scriptstyle I}]=$$-1.13\pm0.16$ ($\sim$0.07 Z$_{\odot}$) toward RBS 144 \citep{Fox13}, $\rm [O\ {\scriptstyle I}/H\ {\scriptstyle I}]=$$-1.24\pm0.20$ ($\sim$0.06 Z$_{\odot}$) toward NGC7714 \citep{Fox13}, and $\rm [O\ {\scriptstyle I}/H\ {\scriptstyle I}]=$$-1.00\pm0.13$ ($\sim$0.10 Z$_{\odot}$) toward NGC7469 \citep{Fox10}. Therefore, the metallicity of the VHVC is at least 0.56 dex (3.3$\sigma$) higher than the mean metallicity of the Magellanic Stream. In particular, it is 0.68 dex (4$\sigma$) higher than the metallicity measured at the tip of the Magellanic Stream toward NGC 7714, which is the closest point to the VHVC in projection (see Fig \ref{fig4}).

We also compare the ionic column densities of the VHVC with those of the nearby ionized gas that is presumably associated with the Magellanic Stream. The top panel of Fig \ref{fig5} shows the $v_{\rm LSR}$ velocities of Magellanic-related ionized absorbers found in \citetalias{Fox14}. This velocity map shows that the ionized gas, regardless of their origins, move at similar velocities in this region. The potential diversity of gas origins can be seen in the middle and bottom panels. In the middle panel, we show the log N(\SiIII) values for the ionized gas; the majority of the Magellanic ionized stream near the tip has log N$_{\rm SiIII}\lesssim$13.0 dex, which is $\sim$0.3 dex lower than that of the VHVC (log N$_{\rm SiIII}\sim13.3$ dex). This difference is shown more obviously when we plot the log N(\SiIII) values against Magellanic latitude. The VHVC's column density is $\gtrsim$0.5 dex higher than those of the Magellanic Stream at similar latitude. Similar trends can also be found in \SiII\ and \CII. The ionized envelope of the Magellanic Stream tends to become more and more diffuse at further distances from its \HI-bright main body. The VHVC, with higher ionic column densities, is unlikely to be associated with the Stream.

We conclude that \textsl{the VHVC is unlikely to originate from the Magellanic Stream}. Nearby our VHVC, there is an \HI-VHVC -- Wright's Cloud -- moving at a similar Magellanic velocity. Some authors have suggested Wright's Cloud may be related to the Magellanic Stream due to their proximity in position-velocity space (\citetalias{Wright79}; \citealt{Braun04, Putman09, Nidever10}). Indeed, \citetalias{Fox14} included Wright's Cloud as one of the \HI-bright regions of the Magellanic System, thus extended the Magellanic ionized envelope 30$\degree$ beyond Wright's Cloud (see Fig 1 in \citetalias{Fox14}). However, as we discuss in Section \ref{sec5.5}, Wright's Cloud's origin is unclear due to its unknown distance. Therefore, we cast doubt on the existence of the Magellanic ionized envelope within 30$\degree$ of Wright's Cloud where there is no other Magellanic-associated \HI\ emission or QSO absorbers. The highly uncertain origin of Wright’s Cloud  and the evidence that the VHVC is not part of the Stream together suggest that the ionized envelope of the Magellanic Stream may not extend to cover the region of the sky in the direction of M33.

\section{The Origin of the VHVC}
\label{sec5}

In this section we discuss the multiple possibilities for the origin of the VHVC given the complex velocity field on this projected sky. Apart from the Magellanic scenario (Case B) that we argue against in the previous section, there are four other plausible scenarios that could cause the absorption, which we illustrate in Fig \ref{fig1} as: A -- a MW CGM cloud, C -- an M31 CGM cloud, D -- M33's northern warp, and E -- M33's outflow. Here we evaluate each of these possibilities separately. 

\begin{figure}[b]
\begin{center}
\includegraphics[width=\columnwidth]{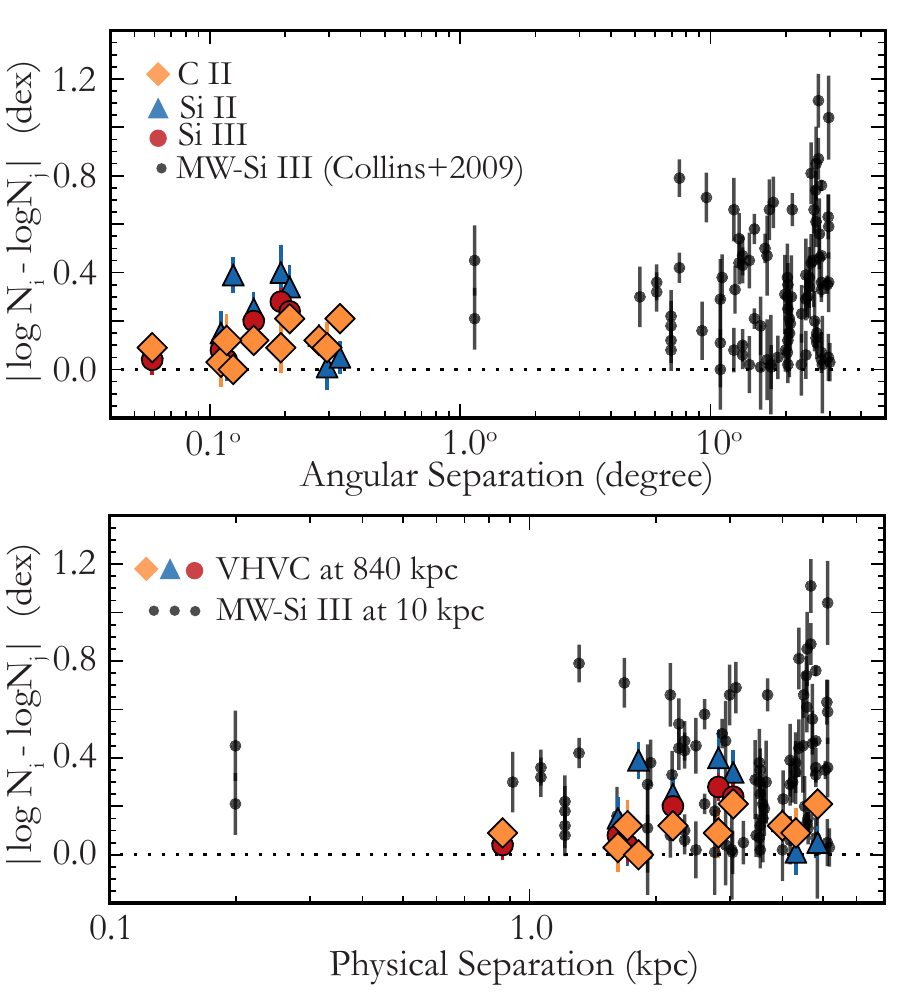}
\caption{The log-log angular variation of the ion column densities. The orange diamonds, blue triangles, and red circles are respectively the \CII, \SiII, and \SiIII\ logarithmic column density differences of the VHVC along our sightline pairs; same as those in Fig \ref{fig3}. The black dots are the \SiIII\ logarithmic column density differences of MW HVCs measured toward sightline pairs with angular separations less than $30\degree$ \citep{Collins09}.  Data points without error bars mean the errors are smaller than the sizes of the symbols.  \label{fig6}%
}
\end{center}
\end{figure}

\subsection{A MW CGM Cloud?}
\label{sec5.1}

As we annotate as Case A in Fig \ref{fig1}, the VHVC could be a cloud at an unknown distance in MW's CGM. Though we have shown in Section \ref{sec4} that the VHVC is unlikely to be part of the Stream, it could be closer in the MW's inner halo similar to other MW HVCs. The $\sim$0.3 Z$_{\odot}$ metallicity of the VHVC makes it similar to some \HI\ complexes, such as Complex C (Z$=$0.1$-$0.3 Z$_{\odot}$; \citealt{Tripp03, Shull11}). In Fig \ref{fig6}, we compare the column density (log N) variations of the VHVC with those MW HVCs detected in \SiIII\ between 58 QSO sightlines with angular separation of $\leq$30$\degree$ \citep{Collins09}. In the top panel, we find that the VHVC's log N values vary up to $\sim$0.4 dex over 0.4$\degree$, while those of the MW HVCs are $\sim$3 times larger. The larger variations of the MW HVCs still hold when we separate Collins et al.'s sample by positive/negative HVC velocities and by northern/southern QSO sightlines.

We note that the direct comparion in angular seperation between our VHVC and the MW HVCs may not be informative since it does not reflect the actual physical separations of the paired sightlines on the sky. The MW HVCs are generally at distance of 5$-$15 kpc from the Sun \citep{Wakker01, Wakker08, Lehner12}; we assume them at $d=$10 kpc and find physical separations of $\lesssim5$ kpc for these MW paired sightlines as shown in the bottom panel of Fig \ref{fig6}. In order to make our VHVC's physical scale comparable to those of the MW HVCs, we find its distance would have to be $\sim$840 kpc, which is beyond the virial radius of the Galaxy and is as far as M33. Therefore, our sightlines toward the VHVC should probe a smoother medium on sub-degree scales than those in \cite{Collins09} if it belongs to the MW's CGM. However, if the VHVC is instead associated with M33 as we discuss in Section \ref{sec5.3} and \ref{sec5.4}, the smoother column density variation in the bottom panel suggests that the VHVC is unlikely to represent HVC-like features in M33's CGM, assuming that M33 has similar HVC population as MW.

\subsection{An M31 CGM Cloud?}
\label{sec5.2}

The VHVC moves at $v_{\rm LSR}\sim-350$ \kms\ while M31's systemic velocity is $v_{\rm LSR}=-301$ \kms\ \citep{Courteau99}. If put at M31's distance, the VHVC would be $\sim$200 kpc from the galaxy in projection. We show this possibility as Case C in Fig \ref{fig1}. By comparing the ion column densities of the VHVC with those detected in M31's CGM by \cite{Lehner15}, we find that the VHVC generally has higher \CII, \SiII, and \SiIII\ column densities. Beyond 50 kpc of M31's CGM, \cite{Lehner15} reported no significant detection of \SiII\ with an upper limits of log N(\SiII)$\lesssim$13.19. In addition, \SiIII\ absorption lines were detected along four sightlines with $<$log N(\SiIII)$>=$12.53, while \CII\ column densities are mostly upper limits (log N(\CII)$<$13.70) with only one detection at log N(\CII)$=$13.16. These measurements indicate that the ionized gas in the outskirt of M31's CGM is most likely to be diffuse, with column densities lower than those we measured for the VHVC: $<$log N(\SiII)$>=$13.29, $<$log N(\SiIII)$>=$13.27, and $<$log N(\CII)$>=$14.13. Given the column density mismatch, we disfavor an origin of the VHVC being part of M31's CGM. Note that M31's CGM may be patchy, similar to the cool CGM of L$^*$ galaxies in the COS-Halos survey \citep{Werk13}; however, the chance of our sightlines intercepting a very-high density clump at large distance from the host galaxy is  rare. Future work on the patchiness of M31's CGM and those of L$^*$ galaxies may help to confirm or dispute our argument for no association between the VHVC and M31.

\subsection{An M33 CGM Cloud or an M33 Outflow?}
\label{sec5.3}

M33 has a systemic velocity of v$_{\rm LSR}\sim-180$ km s$^{-1}$ \citep{Corbelli97}, therefore the VHVC would be moving at $\delta v\sim-170$ km s$^{-1}$ in the galaxy's rest frame. If we simplify M33's gravitational potential as a point mass $\sim$5.5$\times$10$^{10}$ M$_{\odot}$ (dark matter+baryon; \citealt{Corbelli03}), the VHVC should be within $\sim$16 kpc of the galaxy to remain gravitationally bound. This calculation certainly omits the complex mechanisms (ionization, cooling, equilibrium, etc.) that regulate clouds' survivability in a galaxy's CGM, however, it hints that the VHVC is most likely to be near M33's disk if it has some M33 origins. Observationally, the COS-Halos survey \citep{Werk13} found that most of the CGM clouds of L$\sim$L$^*$ galaxies at z$\sim$0.2 were within $\sim$100 km s$^{-1}$ of their host galaxies' systemic velocities and only a few exceed 200 km s$^{-1}$. Therefore, it is unlikely that our VHVC is a CGM cloud of M33 at large radius given its high velocity. As for the possibility that the VHVC exists in the vicinity of M33's disk, we have shown in Section \ref{sec5.1} that the VHVC is unlikely to be HVC-like features due to its smooth column density distribution. In the following, we consider two other near-the-disk scenarios: (1) the VHVC represents an outflow from M33 (Case E in Fig \ref{fig1}), which we discuss in the rest of this section; (2) the VHVC could be an ionized extension of M33's northern warp (Case D in Fig \ref{fig1}; see Section \ref{sec5.4}).

The VHVC should lie in front of M33's disk since our background targets are UV-bright stars in M33. If we use these M33 stars as the reference points, the VHVC is moving at $\sim$[$-$180,$-$100] \kms. These negative velocities with respect to M33 indicate that the VHVC could be regional outflows from the disk but remain in the vicinity of the disk; the outflows could be driven by stellar winds or supernova feedback (Case E in Fig \ref{fig1}) -- Outflows powered by an active galactic nuclei are not considered since M33 does not host a massive central black hole \citep{Merritt01}. In the following, we evaluate the possibility of outflow by comparing the ion properties of the VHVC with those of galactic outflows seen in absorption in other galaxies.

First, we find that our COS spectra do not show a detection of \FeII\ -- an ion that is commonly reported in galactic outflow observations of both nearby (e.g., LMC; \citealt{Lehner09}) and $z$$\sim$1 star-forming galaxies \citep{Martin12, Rubin10, Rubin14}. The non-detection is unlikely to be due to the sensitivity limit of our COS spectra, which is sensitive to \FeII\ absorption lines with log N(\FeII)$\gtrsim$13.5. If M33 indeed has an outflow similar to those in nearby and $z$$\sim$1 galaxies (e.g., \citealt{Martin12, Rubin14}), the \FeII\ column densities should be log N(FeII)$\gtrsim$14.0, well above our sensitivity limit.

Secondly, we investigate the possibility based on ion absorption line widths and profile shapes. Galactic outflows have been detected in several different ions, such as \NaI, \MgII, and \FeII\ with broad absorption lines (FWHM$\gtrsim$100 \kms; \citealt{Heckman00, Martin05,Martin12, Weiner09, Chen10, Rubin10, Rubin14}). Recently, \citet[][hereafter C16]{Chisholm16} studied \SiII/\SiIII-tracing outflows in 37 nearby star-forming galaxies observed with {\it HST}/COS G130M. They generalized that galactic outflows detected in \SiII\ and \SiIII\ commonly have FWHM values of $\sim$230$-$500 \kms\ and the outflows' line profiles tend to exhibit more extended, gradually shallowing blue absorption edges compared to the red edges. This profile asymmetry may be explained by outflows being continuously accelerated with more diffuse gas columns at larger distances. Since our spectra were observed with the same instrument, we can directly compare our \SiII/\SiIII\ measurements with \citetalias{Chisholm16}'s. As we show in Section \ref{sec3.2}, except S2 that is blended with absorption from M33, the absorption lines of the VHVC have a mean FWHM $\lesssim$70 \kms, which is at least a factor of 3 narrower than those of galactic outflows reported by \citetalias{Chisholm16}. In addition, our absorption line profiles are mostly symmetric relative to the line centroids as are shown in Fig \ref{fig2}. We note that the 37 galaxies in \citetalias{Chisholm16} span distances of 3$-$880 Mpc, corresponding to COS aperture physical sizes of 0.04$-$11 kpc. Within this dynamical range \citetalias{Chisholm16} found no correlation between the COS aperture physical sizes and galactic wind line widths. Therefore, although the physical COS aperture size in our observation ($\sim$0.01 kpc) is smaller, this difference is not likely to cause the distinct line widths and line profile shapes. We conclude that the VHVC is unlikely to represent an outflow from M33's disk.

\begin{figure}[t]
\begin{center}
\includegraphics[width=0.95\columnwidth]{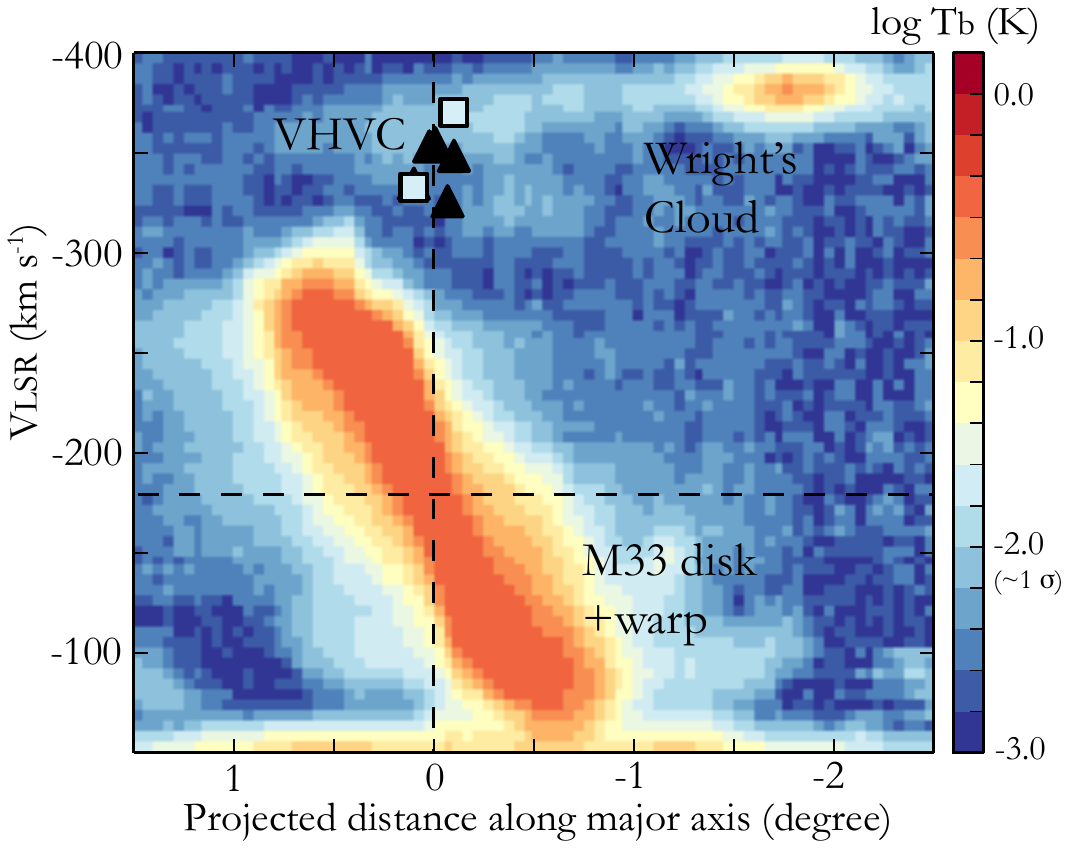}
\caption{Position-velocity map projected along the major axis of M33. The color represents the mean \HI\ brightness temperature on a logarithmic scale (log T$_b$); the 1$\sigma$ value of T$_b$ is $\sim$0.008 K \citep{Auld06}. The major axis is defined based on M33's central optical disk (PA$\sim$21$\degree$; \citealt{Corbelli97}); we show the relative positions of M33's disk, warp and its major axis in Fig \ref{fig8}. X axis shows the angular distance along the major axis, with positive values to the north-east. The bottom emission is the \HI\ from the MW's ISM. The elongated orange-red feature dominating the center of the figure is M33's disk, and the diffuse, light-blue emission in its surroundings shows the extended warp of M33 (\citealt{Wright74}; \citetalias{Wright79}). The VHVC is indicated by light-blue squares (\HI\ emission) and black triangles (\SiII\ absorption) with other ions at a similar position-velocity. The colors of the blue squares indicate the peak brightness temperature ($\approx$0.015 K) of the VHVC detected along S2 and S7 (see Table \ref{tb1}). Note that this figure makes it appear as if there is an \HI\ bridge-like feature between the VHVC and Wright's Cloud. This is a diffuse \HI\ cloud identified by Keenan et al. (\citeyear{Keenan16}, numbered as AGESM33-31; see also \citealt{Grossi08}), which we annotate as``K16 cloud" in Fig \ref{fig8}. The K16 Cloud is on the opposite side of M33 (see Fig \ref{fig8}) and therefore does not have a physical spatial connection to Wright's Cloud. \label{fig7}}
\end{center}
\end{figure}

\subsection{M33's Extended Warp Material?}
\label{sec5.4}

The northern warp of M33 moves at negative velocities of $v_{\rm LSR}$$\sim$[$-$300, $-$200] \kms\ \citep{Putman09, Corbelli97}, which is close to the velocity of the VHVC. Although the northern warp's orientation is unclear, it is possible that the warp folds toward the MW and cause the ionized gas absorption along our line of sight (see Case D in Fig \ref{fig1}). That M33's northern warp may have a diffuse extension is hinted by the deeper \HI\ maps shown in Putman et al. (\citeyear{Putman09}; Fig 3 \& 7) and in Keenan et al. (\citeyear{Keenan16}; Fig 1), which both detected some diffuse \HI\ emission at $v_{\rm LSR}\lesssim-250$ km s$^{-1}$ near the warp. In particular, \cite{Putman09} suggested the diffuse \HI\ structures cannot be reproduced in M33's tilted-ring model \citep{Corbelli97} and an additional component is in need to explain these features.

To examine the possible connection between the VHVC and the northern warp, in Fig \ref{fig7} we show the position-velocity map of M33's \HI\ disk and warp projected along its major axis. The major axis is defined based on M33's central optical disk with a position angle (PA) $\sim$21$\degree$ as adopted in \cite{Corbelli97}. We show the relative positions of M33's warp and central disk in contours in Fig \ref{fig8}, along with a straight line indicating its major axis. To generate the position-velocity map in Fig \ref{fig7}, we project all the \HI\ signals within $26\degree>$RA$>21\degree$ and $28.6\degree<$Dec$<32\degree$ along M33's major axis and compute the mean T$_b$. From Fig \ref{fig7}, we find that the VHVC could be potentially associated with M33's northern warp as revealed by their proximity in the position-velocity space.

If the VHVC is part of M33's northern warp, then the warp folds toward the MW and is stable as inferred from the smooth column density and velocity distributions of the VHVC (Fig \ref{fig6}). In addition, the VHVC's metallicity of $\rm [O\ {\scriptstyle I}/H\ {\scriptstyle I}]=$$-$0.56$\pm$0.17 dex implies that the warp has a similar oxygen abundance to M33's ISM ($\rm [O\ {\scriptstyle I}/H\ {\scriptstyle I}]=$$-$0.42$\pm$0.06 dex; \citealt{Crockett06}). This similarity rules out the possibility that M33's warp represents the accretion of primordial \HI\ from outside of the disk. Instead, it favors the scenario that M33 closely interacted with M31 in the past which resulted in a distorted gaseous warp \citep{Putman09} and stellar disk \citep{McConnachie09}. This contradicts a recent dynamical study of M33's orbit history \citep{Patel16} that suggests a rare chance ($<$1\% at 4$\sigma$) of an M33-M31 close encounter in the past.

\subsection{Wright's Cloud Association?}
\label{sec5.5}

\begin{figure}[t]
\begin{center}
\includegraphics[width=\columnwidth]{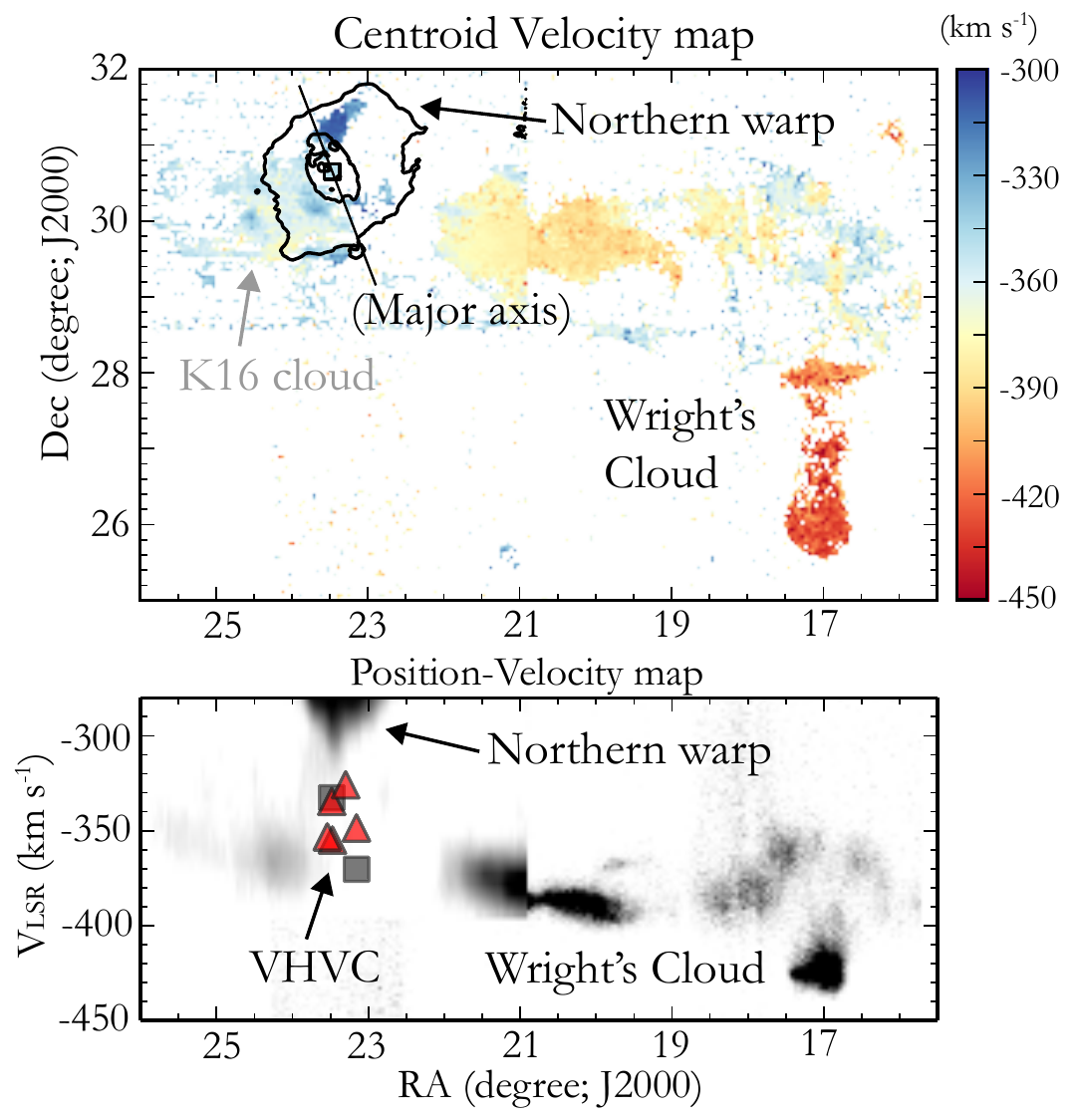}
\caption{Top: the velocity distribution of the VHVC, Wright's Cloud, and M33's northern warp. The map includes AGES \HI\ data (DEC$>$28.5, RA$>$21) and GALFA-HI DR2 data (the rest). The AGES (GALFA-HI DR2) map is shown with log N(\HI)$\geq$18.0 (log N(\HI)$\geq$19.0) for better illustration. The maps are integrated from $v_{\rm LSR}=$$-$450 to $-$300 \kms, therefore M33's velocity field ($-$300$\lesssim$v$_{\rm LSR}$$\lesssim$$-$50 \kms; see Fig \ref{fig7}) is not shown here.  Instead, we overlay two \HI\ column-density contours to show the positions of M33's warp (outer contour; log N$=$19.3) and central disk (inner counter; log N$=$21.0). In addition, we show M33's major axis, which is used in producing the position-velocity map in Fig \ref{fig7}. The VHVC is noted as a small square near M33's disk center. Bottom: position-velocity map. The VHVC is indicated by its \HI\ emission (grey squares) and \SiII\ absorption (red triangles). There exists a discontinuity at RA$\sim21\degree$ which is artificial due to the different column density levels we adopted for AGES and GALFA-HI data.  \label{fig8}% 
}
\end{center}
\end{figure}

Wright's Cloud is composed of north-south and east-west arms as shown in Fig \ref{fig8}. It is $\sim$2$\degree$ from our VHVC and moves at $v_{\rm LSR}\sim$[$-$450, $-$330] \kms, similar to our VHVC. These two clouds' similar velocities and close angular distance suggest a potential connection. In the bottom position-velocity panel, Wright's Cloud can be found with several noticeable \HI\ concentrations at RA$\sim$17$-$22$\degree$. The VHVC is represented by its \HI\ emission and \SiII\ absorption at RA$\sim$23.5$\degree$, generally following the velocity gradient of Wright's Cloud. 

The distance to Wright's Cloud is unknown. Some authors have indicated that Wright's Cloud could be part of the Magellanic Stream based on position-velocity proximity (e.g., \citetalias{Wright79}; \citealt{Braun04, Putman09, Nidever10}). However, if the VHVC is related to Wright's Cloud, its metallicity of Z$=$0.28 Z$_{\odot}$ (Section \ref{sec3.1}) implies that Wright's Cloud, together with the VHVC, is unlikely to be related to the Magellanic Stream. 

If the VHVC is part of Wright's Cloud, then Wright's Cloud could be a nearby \HI\ feature in the MW's inner halo (see Section \ref{sec5.1}). On the other hand, if we combine Fig \ref{fig7} and \ref{fig8}, there is a possibility that our VHVC is forming an ionized bridge that connects Wright's Cloud and M33's northern warp. If Wright's Cloud is in M33's CGM, it would contain a total mass of M(\HI)$\approx$4.8$\times$10$^8$ M$_{\odot}$(d/ 840 kpc)$^2$ as calculated using GALFA-\HI\ data (see also \citetalias{Wright79}, ;\citealt{Braun04}). Wright's Cloud would be $\sim$70 kpc wide in size and $\sim$10 kpc at its nearest point to M33 if put at the galaxy's distance, as have been pointed out in \citetalias{Wright79}. In such case, Wright's Cloud would be a Magellanic-like \HI\ complex in M33's CGM. 

With an estimated \HI\ mass of $\sim$10$^8$ M$_{\odot}$ at the distance of M33, Wright's Cloud may then have an optical counterpart. However, there is no obvious optical counterpart in Wright's Cloud in the Pan-STARRS1 database \citep{Bernard16, Chambers16}. In addition, we briefly search the GALEX map and Planck dust map \citep{Planck15} on the MAST archive but do not find any UV or dust emission excess. If Wright's Cloud does not have an optical counterpart,  it is an outlier from those \HI\ sources (M$_{\rm HI}$$\gtrsim$10$^7$ M$_{\odot}$) found in \cite{Pisano07, Pisano12} which have stellar components.  It is more likely to be similar to the Magellanic Stream or other \HI\ features surrounding galaxies as noted by Sancisi et al. (\citeyear{Sancisi08}; and references therein).

\section{Conclusion}
\label{sec6}

We report the detection of a VHVC in direction of M33. The detection was made along five {\it HST}/COS sightlines targeted at UV-bright stars in M33's disk. The VHVC was found in \OI, \CII, \SiII, and \SiIII\ absorption lines at $v_{\rm LSR}\sim$$-$350$\pm$15 \kms. We do not find \FeII, \PII, \SII, or \SiIV\ absorption at similar velocities even though our COS spectra cover these ionic species. The mean ion column densities of the VHVC are log N(\OI)$=$14.00$\pm$0.07, $<$log N(\CII)$>$$=$14.13$\pm$0.08, $<$log N(\SiII)$>$$=$13.29$\pm$0.15, and $<$log N(\SiIII)$>$$=$13.27$\pm$0.11. 

We measure a metallicity [\OI/\HI]$=$$-$0.56$\pm$0.17 (Z$=$0.28$\pm$0.11 Z$_{\odot}$) for the VHVC, which is 0.68 dex (4$\sigma$) higher than the metallicity found for the tip of the Magellanic Stream. The higher metallicity and stronger \CII, \SiII, and \SiIII\ absorption suggest that the VHVC is most likely not associated with the Magellanic Stream. Furthermore, this lack of association may signify that the ionized envelope of the Magellanic Stream is less extended than previously reported in the direction toward M33.

We find that the VHVC is unlikely to be related to M31's CGM since it would be $\sim$200 kpc from M31's disk in projection but its ionic lines are much stronger than those detected beyond 50 kpc of M31's disk. We also rule out the possibility that the VHVC resides in M33's CGM at large radius given its high velocity ($\delta v\sim$$-$170 km s$^{-1}$) with respect to M33's systemic velocity. If the VHVC has some M33 origins, it has to be within $\sim16$ kpc from the galaxy to remain gravitationally bound. In the vicinity of M33's disk, we find that the VHVC is unlikely to represent an outflow from M33. This is because the ionic lines of the VHVC are narrower and more symmetric than those detected in galactic outflows seen down the barrel of nearby galaxies.

There remain three intriguing possibilities highlighted by our analysis. First, the VHVC could be a normal ionized absorber sitting in MW's CGM. Second, it could be associated with the nearby Wright's Cloud, which would indiate that Wright's Cloud does not belong to the Magellanic Stream. Third, it could be part of M33's northern warp given the similar metallicity and proximity in position-velocity space. If true, the data indicate that M33's northern warp is folding toward the MW. The VHVC's metallicity would imply an ISM origin for the warp, favoring the scenario that M33's warp was formed during a past interaction between M33 and M31. Furthermore, the proximity of the VHVC with Wright's Cloud and M33's northern warp may hint that these three objects are physically connected. In this case, Wright's Cloud would be a Magellanic-like structure in M33's CGM, which would have important implications on the dynamical history of M33. To break the degeneracy of all these scenarios, further deep \HI\ observations to map diffuse \HI\ gas in this Magellanic-M33-M31 overlapping area would be highly valuable. Absorption-line experiments using MW halo stars toward this direction would also help to bracket (or yield lower/upper limits on) the distance.

{\it Acknowledgement.} We thank the referee for the useful comments in improving our manuscript. We thank R. F. Minchin and the AGES team for their generosity in providing us the high-sensitivity \HI\ data of M33. We thank A. J. Fox and the COS team for their help in providing information about the night-only data reduction. We appreciate A. J. Fox's great feedback on Section \ref{sec4} where we discuss the origin of the VHVC with respect to the Magellanic Stream. This work is based on observations made with the NASA/ESA Hubble Space Telescope (program ID: 13706). Support for HST-GO-13706 was provided by NASA through a grant from the Space Telescope Science Institute (STScI). STScI is operated by the Association of Universities for Research in Astronomy, Inc., under NASA contract NAS5-26555. Some of the data presented in this paper were obtained from the Mikulski Archive for Space Telescopes (MAST).  We also acknowledge support from the National Science Foundation under Grant No. AST-1410800. J.E.G. Peek and J.K. Werk were supported in part by Hubble Fellowship grants 51295 and 51332, respectively, provided by NASA to STScI. This research made use of the IPython package \citep{Ipython07}, matplotlib \citep{Hunter07} which is a Python library for publication quality graphics, and Astropy \citep{Astropy13} which is a community-developed core Python package for Astronomy. CalCOS is a product of the Space Telescope Science Institute, which is operated by AURA for NASA. 

\bibliographystyle{apj}
\bibliography{Zheng17_iVHVC}

\end{document}